\documentclass[conference]{IEEEtran}
\usepackage{color}
\usepackage{lettrine}
\usepackage{multirow}
\usepackage{algorithm,algorithmic}
\newcommand{\rpm}{\sbox0{$1$}\sbox2{$\scriptstyle\pm$}
\raise\dimexpr(\ht0-\ht2)/2\relax\box2 }
\usepackage{graphicx}
\usepackage{booktabs}
\usepackage{adjustbox}
\usepackage{tabularx}
\usepackage{cite}
\usepackage{blindtext}
\usepackage{authblk}
\usepackage{amsmath}
\usepackage{comment}
\usepackage{pifont}
\newcommand{\tick}{\ding{51}}
\usepackage{eso-pic}

\begin{document}
\AddToShipoutPicture*{\small \sffamily\raisebox{1.2cm}{\hspace{1.8cm}978-1-7281-0397-6/19/\$31.00 ©2019 IEEE}}
\title{Variation-aware Binarized Memristive Networks}
\author[1]{Corey Lammie}
\author[2]{Olga Krestinskaya}
\author[3]{Alex James}
\author[1]{Mostafa Rahimi Azghadi}

\affil[1]{College of Science and Engineering, James Cook University, Queensland 4814, Australia \authorcr Email:\{corey.lammie, mostafa.rahimiazghadi\}@jcu.edu.au}
\affil[2]{School of Engineering, Nazarbayev University, Kazakhstan, Email: okrestinskaya@nu.edu.kz}
\affil[3]{AI division, Clootrack Pvt Ltd, Bangalore, India,  Email: apj@ieee.org}

\maketitle

\begin{abstract}
The quantization of weights to binary states in Deep Neural Networks (DNNs) can replace resource-hungry multiply accumulate operations with simple accumulations. Such Binarized Neural Networks (BNNs) exhibit greatly reduced resource and power requirements. In addition, memristors have been shown as promising synaptic weight elements in DNNs. In this paper, we propose and simulate novel Binarized Memristive Convolutional Neural Network (BMCNN) architectures employing hybrid weight and parameter representations. We train the proposed architectures offline and then map the trained parameters to our binarized memristive devices for inference. To take into account the variations in memristive devices, and to study their effect on the performance, we introduce variations in $R_{\textnormal{ON}}$ and $R_{\textnormal{OFF}}$. Moreover, we introduce means to mitigate the adverse effect of memristive variations in our proposed networks. Finally, we benchmark our BMCNNs and variation-aware BMCNNs using the MNIST dataset.
\end{abstract}

\IEEEpeerreviewmaketitle
\section{Introduction}

\begin{figure*}[!t]
	\centering
	\includegraphics[width=1\textwidth]{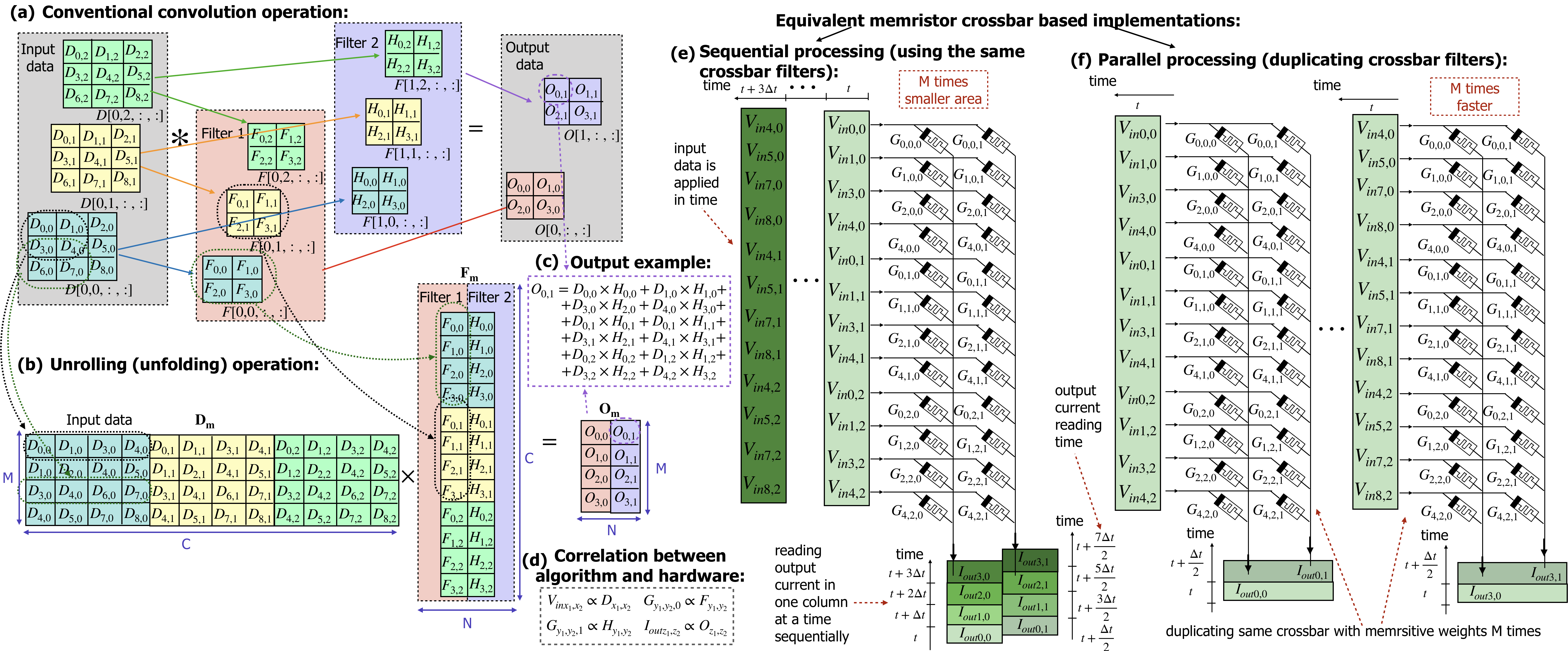}
	\caption{Depiction of (a) the reduction of a conventional convolutional layer to (b) an unrolled matrix multiplication, (c) an output example, (d) the correlation between the algorithm and hardware, and corresponding hardware implementation of convolutional layers for (e) sequential and (f) parallel processing.}
	\label{im2col}
\end{figure*}

\lettrine{R}{esistive} Random Access Memory (ReRAM) is a class of memristors, that when arranged in a crossbar configuration, can be used to implement multiply and accumulate (MAC) or dot-product multiplications consuming low energy and area on chip. ReRAM devices, in such configurations, can be used to reduce the time complexity of 2D matrix-vector multiplications, used extensively during forward and backward propagation cycles in DNNs, from $\mathcal{O}(n^2)$ to $\mathcal{O}(n)$, and in extreme cases to $\mathcal{O}(1)$. However, current ReRAM crossbars face concerns of aging, non-idealities and endurance~\cite{8702245}, that limit the accuracy of their conductive states, affecting the reliability and robustness of memristive DNNs.

Memristive DNNs can either employ ReRAM crossbars with multiple distinctive conductive states, to represent analog weight representations, or with two distinctive conductive states, to represent binary weight states. Given the aging and state variability issues of ReRAM, binary weight representations, adopted in BNNs, are currently more practical for hardware realization.

Binarized Neural Networks (BNNs)~\cite{CourbariauxB16}, which perform binary MAC computations during forward and backward propagations, have demonstrated comparable performance to conventional DNNs, while significantly reducing resource and power utilizations~\cite{DBLP:journals/corr/abs-1905-06105}. On account of endurance concerns, ReRAM devices are ill-suited for implementing backward propagations, required during the training routine of BNNs where a large number of programming cycles are required. However, they are well-suited~\cite{DBLP:journals/corr/abs-1808-00737} for implementing forward propagations, required during inference, as only a limited number of programming cycles and two conductive states are required.

In this paper, we propose and simulate novel BMCNNs and variation-aware BMCNNs using a customized simulation framework for memristive crossbars, which integrates directly with the $PyTorch$ Machine Learning (ML) library. Our developed networks employ offline training routines adopting hybrid fixed-point and floating-point representations, and binarized memristive weights. Furthermore, to reduce the effect of memristor variability on the performance of our architectures after crossbar programming, we propose a tuning method.
The specific contributions of this work are as follows:

\begin{itemize}
  \item We propose and simulate novel BMCNNs and variation-aware BMCNNs, adopting hybrid fixed-point, floating-point, and binarized parameter representations, simulating memristive devices, and benchmark them using the MNIST dataset.
  \item We investigate the performance degradation observed when the variance of $R_{\textnormal{ON}}$ and $R_{\textnormal{OFF}}$ are increased within memristive crossbars that compute matrix multiplication operations for convolutional layers during inference.
  \item We propose a tuning method to reduce the effects of memristor variability without reprogramming memristive devices.
\end{itemize}

\section{Preliminaries}
This section briefly reviews and presents the algorithms and methods used in our developed architectures.

\subsection{Binary Weight Regularization}
Binary weight regularization~\cite{CourbariauxB16}, constrains network weights to binary states of \{+1, -1\} during forward and backward propagations. The binarization operation transforms full-precision weights into binary values using the signum function, described in Eq. (\ref{det_binarization}).

\begin{equation}\label{det_binarization}
w_b = \textnormal{sign}(w) = \left\{\begin{array}{lr}
-1 & \textnormal{if } w \leq 0\\
+1 & \textnormal{otherwise},
\end{array}\right.
\end{equation}
where $w_b$ is the binarized weight and $w$ is the full-precision weight. During backward propagations, large weights are clipped using $t_{\textnormal{clip}}$, described in Eq. (\ref{STE}), where $c$ denotes the objective function.

\begin{equation}\label{STE}
  \frac{\partial c}{\partial w} = \frac{\partial c}{\partial w_b} 1_{|w| \leq t_{\textnormal{clip}}}
\end{equation}

\subsection{Convolutional Operation as a Matrix Multiplication}
Convolutional operations in BMCNNs can be performed using unrolling techniques, which reduce conventional convolution operations to matrix multiplications. Fig. \ref{im2col} (a) and (b) depict the computation of the convolution of two filters $(f = 2)$ using conventional and unrolling techniques. In Fig. \ref{im2col} (b), both convolutional filters, $\mathbf{F}$ and $\mathbf{H}$, are reshaped to form $\mathbf{F_m}$ of size $(C \times N)$. The input, $\mathbf{D}$, is reshaped to form $\mathbf{D_m}$ of size $(M \times C)$. The convolution result, $\mathbf{O}$, is determined by reshaping the result of $\mathbf{F_m} \times \mathbf{D_m}$, $\mathbf{O_m}$, from $(M \times N)$ to $(f \times o_1 \times o_2)$, where $o_1 = o_2 = ([i_2 - k_2 + 2 \times P] / S) + 1$, which in this instance is 2.

Fig. \ref{im2col} (e) and (f) depict the mapping of the matrix multiplication operation to memristive crossbars using sequential and parallel processing approaches. Elements of $\mathbf{D_m}$ are represented using equivalent voltages, $\mathbf{V_{in}}$. In the first column of the crossbar, memristors are programmed to $G_{y_1,y_2,0} \propto F_{y_1,y_2}$. In the second column of the crossbar, memristors are programmed to $G_{y_1,y_2,1}\propto H_{y_1,y_2}$ (see Fig. \ref{im2col} (d)). The total output current from the crossbar, $\mathbf{I_{out}}$, is lineary proportional to the convolution result, i.e. $\mathbf{O}	\propto K \cdot \mathbf{I_{out}}$, and can either be read sequentially column-by-column, or in parallel, to reduce the output current error due to leakage.

Convolutional layers can be processed sequentially, using the same crossbar representative of $\mathbf{F_m}$ to process M input rows of $\mathbf{D_m}$ one by one, or in parallel with $M$ crossbars, using the same memristive filter $\mathbf{F_m}$ duplicated $M$ times (see Fig. \ref{im2col} (e)). Despite being much faster, this parallel approach increases the on-chip area by a factor of $M$.

\begin{table}[!b]
\centering
\caption{Memristive BNN architecture.}
\begin{adjustbox}{width=0.5\textwidth}
\begin{tabular}{lcc}\toprule
\textbf{Layer} & \textbf{Binarized} & \textbf{Memristive} \\
\midrule
Convolutional Layer$^1$. f = 16, $k_2$=$k_3$=2, P=2 & \tick & \tick\\
Convolutional Layer$^1$. f = 32, $k_2$=$k_3$=2, P=2 & \tick & \tick\\
Fully Connected Layer$^1$. N = 1568              & & \\
\bottomrule
\multicolumn{3}{l}{$^1$No biases are used.} \\
\end{tabular}
\end{adjustbox}\label{network_architecture}
\end{table}

\begin{figure*}[!t]
	\centering
	\includegraphics[width=1\textwidth]{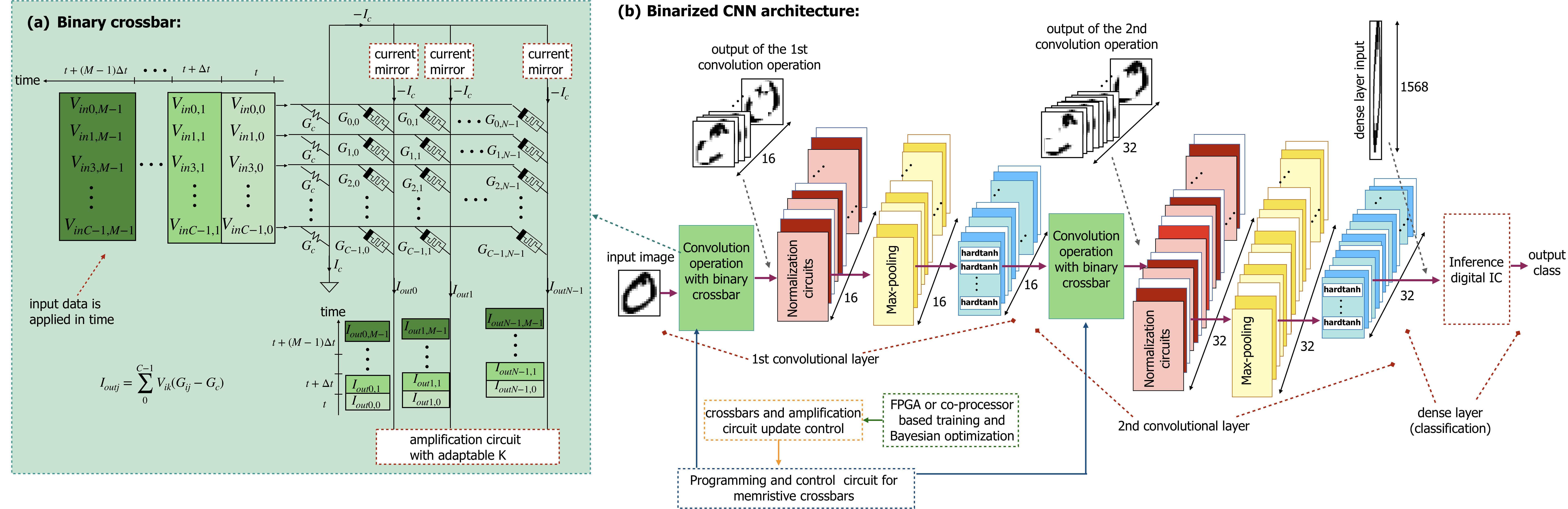}
	\caption{(a) The single-column memristor crossbar array architecture used in all of our FP-8 MBNN and TFP-8 MBNN networks. (b) Overall architecture of our binarized CNNs.}
	\label{crossbar}
\end{figure*}

\section{Network Architecture}
The network architecture adopted by all of our memristive BNNs, originally proposed in~\cite{CourbariauxB16}, is depicted in Fig. \ref{crossbar} (b), and summarized in Table \ref{network_architecture}. All convolutional layers are followed by batch-normalization, max-pooling $(k_2=k_3=S=2)$, and \textit{hardtanh} activation operations. Binary weight representations are used for all convolution layers. We implemented four network architectures, each denoted by a name which includes two parts. The first part denotes the number representation method used for weights during the parameter-update stage, and for the last fully connected layer, while the second part denotes the binary weight representation. For instance, FR BNN describes an architecture that uses (FR) Full-Resolution 32-bit floating point numbers, and (BNN) binarized weights. The other architectures are as follows: 8-bit Fixed-point and Binary (FP-8 BNN), and 8-bit Fixed-point and Memristive Binary (FP-8 MBNN, and TFP-8 MBNN). FP-8 MBNN is used to denote BMCNNs with fixed crossbar current amplification parameters, whereas TFP-8 MBNN is used to denote variation-aware BMCNNs with tuned crossbar current amplification parameters.

Our proposed hardware implementation consists of an offline training module, which can be based on either a FPGA or co-processor, a programming and crossbar control circuit, and forward propagation circuits. These forward propagation circuits utilize several ReRAM crossbars, depicted in Fig. \ref{crossbar} (a), in which each memristors state is confined to $[R_{\textnormal{ON}}$, $R_{\textnormal{OFF}}]$, to represent $[-1, +1]$, respectively~\cite{van2018memristor}. The multiplication of $\mathbf{D_m}\times \mathbf{F_m} = \mathbf{O_m}$, where $\mathbf{D_m}$ contains full resolution elements, and $\mathbf{F_m}$ contains binary elements $\in [-1, +1]$ (see Fig.~\ref{im2col}(d)), can be performed as described in Eq. (\ref{crossbar_math}).

\begin{equation}\label{crossbar_math}
  \begin{aligned}
    \mathbf{O_m}[j, k] &= K \sum_{0}^{C-1} \mathbf{V}_{i,k} (G_{i,j} - G_c)
  \end{aligned}
\end{equation}

Each element in a single row of matrix $\mathbf{O_m}$ is equivalent to the scaled output current from a single crossbar column. Each row of $\mathbf{O_m}$ is computed by applying $\mathbf{V_{in}}$ to each crossbar row in time. To represent both positive and negative binary weights, we introduce a crossbar column with fixed resistors $G_c=[G_{\textnormal{ON}} + G_{\textnormal{OFF}}]/2$, whose current,$-I_c$, is duplicated to all the crossbar columns with memristors using current mirrors~\cite{van2018memristor}. The output current from each column is computed as $I_{\textnormal{out}j,k}=\sum_{0}^{C-1}V_{i,k}(G_{i,j}-G_c)$, where $V_{ini,k}$ is an input to row $i$ at time $k$ and $I_{\textnormal{out}j,k}$ is an output of the column $j$ at time $k$, for $ i = 1 \textnormal{ to }C$, $ j = 1 \textnormal{ to }N$ and $ k = 1 \textnormal{ to }M$.

If the utilized memristive devices are considered ideal, we can pick a current amplification parameter K=4000, which maps devices perfectly to their desired $R_{\textnormal{ON}}$, $R_{\textnormal{OFF}}$ states. To develop a framework for realistic memristors, we perform tuning for our proposed variation-aware BMCNNs to alleviate performance degradation due to memristor variabilities. This tuning process uses Bayesian optimization to determine and set each crossbars adaptable current amplification parameter, $K$, for each layer $\in[3000\textup{:}5000]$ with 15 Bayesian trials.

\section{Implementation Results}

\begin{table}[!t]
\centering
\caption{Tuned network hyperparameters.}
\begin{tabular}{lrllr}\toprule
\textbf{Optimizer} & \textbf{$\Im$} & \textbf{$t_{\textnormal{clip}}$} & \textbf{$\eta$} & \textbf{Validation Set Accuracy} \\
\midrule
\multicolumn{5}{c}{FR-32 BNN} \\
\midrule
AdaGrad                                  &  64 & 0.64 & 1.60E-03 & 97.51 \\
Adam                                     & 128 & 0.91 & 3.78E-03 & 96.45 \\
$\textnormal{SGD}_{\textnormal{m}=0}$    & 119 & 0.56 & 3.79E-03 & 96.71 \\
$\textnormal{SGD}_{\textnormal{m}=0.8}$  & 120 & 0.87 & 1.00E-03 & 97.97 \\
\midrule
\multicolumn{5}{c}{FP-8 BNN} \\
\midrule
AdaGrad                                  &  64   & 1.00  & 4.53E-03  & 94.31 \\
Adam                                     & 128   & 0.68  & 1.00E-03  & 98.35 \\
$\textnormal{SGD}_{\textnormal{m}=0}$    & 128   & 0.50  & 4.61E-03  & 95.97 \\
$\textnormal{SGD}_{\textnormal{m}=0.8}$  & 119   & 0.89  & 3.27E-03  & 96.64
\\ \bottomrule
\end{tabular}\label{tuned_network_parameters}
\end{table}

\begin{figure*}[!t]
	\centering
	\includegraphics[width=0.9\textwidth]{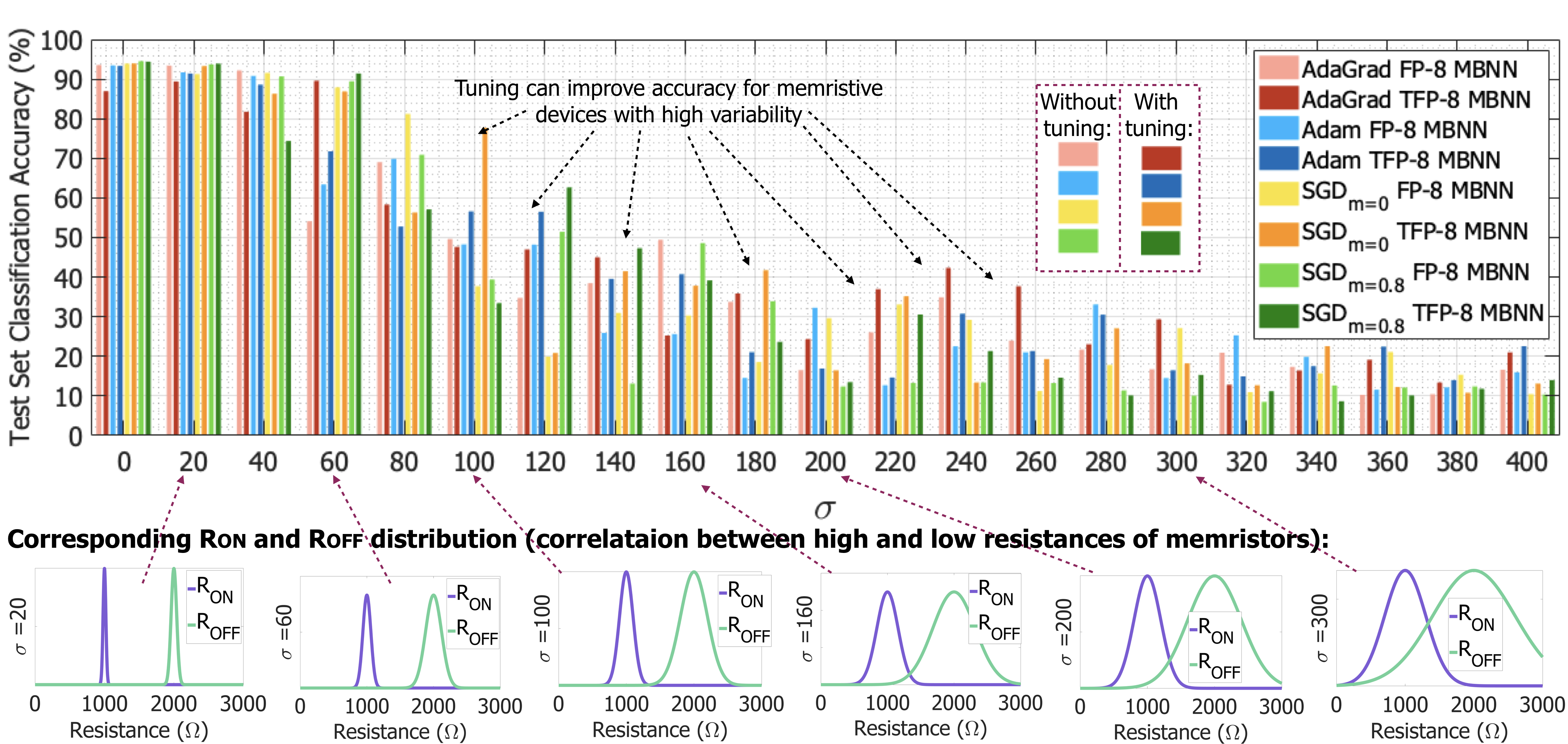}
	\caption{Test set classification accuracy for all FP-8 MBNN and TFP-8 MBNN networks.}
	\label{test_set_plot}
\end{figure*}

In order to investigate the performance of our networks the MNIST dataset was used. During backward propagations Eq. (\ref{STE}) was used to binarize weights, and after the parameter update procedure, their full-precision representations were clipped using $t_{\textnormal{clip}}$.

\subsection{Hyperparameter Optimization}
Prior to training networks using the MNIST training set hyperparameter optimization was performed by constructing modified training and validation sets, using 80\% (48,000) and 20\% (12,000) of training samples, respectively. We performed Bayesian optimization using \textit{Ax} for a batch size $\Im \in [64\textup{:}128]$, $t_{\textnormal{clip}} \in [0.5\textup{:}1.0]$, and a learning rate, $\eta \in [1e^{-3}\textup{:}1e^{-2}]$, for AdaGrad, Adam, $\textnormal{SGD}_{\textnormal{m}=0}$ (SGD with m=0), and $\textnormal{SGD}_{\textnormal{m}=0.8}$.

The best validation set accuracy for each network during 20 training epochs for 15 Bayesian trials is presented in Table \ref{tuned_network_parameters}. We observed no notable drop in the validation set accuracy between our optimized FR-32 BNN and FP-8 BNN implementations. Hence, herein, all memristive BNNs adopt hybrid 8-bit fixed-point and memristive binary weight representations.

\subsection{Memristor Crossbar Programming and Tuning}
After each crossbar was programmed using programming and control circuitry, all trained binarized network weights were programmed to the crossbars and then discarded, requiring no further storage. Tuning was performed for all variation-aware BMCNNs. We note that after training, for our proposed hardware implementation, the offline training module can be freely disconnected.

\subsection{Performance Investigation}
To investigate the performance of our networks, we directly compared the test set classification accuracy of all our BNNs. FP-8 MBNN networks adopted fixed crossbar current amplification parameters, $K=4000$ for each layer, while TFP-8 MBNN networks adopted tunable crossbar current amplification parameters. Simulations of memristive BNNs were performed using a modified Generalized Boundary Condition Memristor $TiO_2$ model~\cite{8351429} with $R_{\textnormal{ON}}=1000\Omega$ and $R_{\textnormal{OFF}}=2000\Omega$. All results are presented in Table \ref{results_table}.

\subsection{Performance Degradation Due to Device Variability}
To determine the effect of memristor variability on the performance of each network, and how the tuning of $K$ improves their accuracy, resistive states for each memristor were sampled from a Gaussian distribution with a standard deviation of $\sigma$ from the trained state of that memristor. As the variability of the $R_{\textnormal{OFF}}$ state can be higher than $R_{\textnormal{ON}}$, we used a larger $\sigma$ value when sampling $R_{\textnormal{OFF}}$, i.e.  $\sigma_{R_{\textnormal{ON}}}=\sigma$ and $\sigma_{R_{\textnormal{OFF}}}=2\sigma$. The performance of all memristive BNNs under such conditions are presented in Fig. \ref{test_set_plot}.

For all networks, a test classification accuracy of $>90\%$ was obtained when $\sigma \leq 40$ and the distributions of $R_{\textnormal{ON}}$ and $R_{\textnormal{OFF}}$ weights did not correlate. Even small correlations of $R_{\textnormal{ON}}$ and $R_{\textnormal{OFF}}$ states caused a substantial drop in accuracy. Fig. \ref{test_set_plot} demonstrates that the proposed tuning method can increase the test set classification accuracy, when $\sigma \geq 100$.

\begin{table}[!t]
\centering
\caption{Comparison of test set classification accuracy (\%) for memristive BNN and digital implementation of BNN with different weight resolutions and optimizations.}
\begin{tabular}{lcccc}
\hline
\textbf{Optimizer} & AdaGrad & Adam & $\textnormal{SGD}_{\textnormal{m}=0}$ & $\textnormal{SGD}_{\textnormal{m}=0.8}$ \\
\hline
FP-8 BNN & 93.68\% & 93.42\% & 93.99\% & 94.31\% \\
FR-32 BNN & 93.93\% & 92.21\% & 94.17\% & 94.11\% \\
\textbf{FP-8 MBNN} & \underline{\textbf{93.56\%}} & \underline{\textbf{93.43\%}} &  93.90\% & \underline{\textbf{94.50\%}}\\
\textbf{TFP-8 MBNN} & 86.94\% & 93.33\% & \underline{\textbf{93.95\%}} & 94.41\%\\
\bottomrule
\end{tabular}
\label{results_table}
\end{table}

\section{Conclusion}
We proposed novel memristive BNNs with tunable crossbar output current amplification factors. We benchmarked the performance of our novel architectures and compared them to the digital implementations of Binarized CNNs. We demonstrated that memristor variabilities can degrade performance, and proposed an alleviating tuning method. We leave the development of full circuit level implementations of the proposed architecture and specific device and technology investigations to future works.

\bibliographystyle{IEEEtran}
\bibliography{SC}

\begin{thebibliography}{1}
\providecommand{\url}[1]{#1}
\csname url@samestyle\endcsname
\providecommand{\newblock}{\relax}
\providecommand{\bibinfo}[2]{#2}
\providecommand{\BIBentrySTDinterwordspacing}{\spaceskip=0pt\relax}
\providecommand{\BIBentryALTinterwordstretchfactor}{4}
\providecommand{\BIBentryALTinterwordspacing}{\spaceskip=\fontdimen2\font plus
\BIBentryALTinterwordstretchfactor\fontdimen3\font minus
  \fontdimen4\font\relax}
\providecommand{\BIBforeignlanguage}[2]{{%
\expandafter\ifx\csname l@#1\endcsname\relax
\typeout{** WARNING: IEEEtran.bst: No hyphenation pattern has been}%
\typeout{** loaded for the language `#1'. Using the pattern for}%
\typeout{** the default language instead.}%
\else
\language=\csname l@#1\endcsname
\fi
#2}}
\providecommand{\BIBdecl}{\relax}
\BIBdecl

\bibitem{8702245}
O.~{Krestinskaya}, A.~{Irmanova}, and A.~P. {James}, ``Memristive
  non-idealities: Is there any practical implications for designing neural
  network chips?'' in \emph{2019 IEEE International Symposium on Circuits and
  Systems (ISCAS)}, May 2019, pp. 1--5.

\bibitem{CourbariauxB16}
I.~Hubara, M.~Courbariaux, D.~Soudry, R.~El-Yaniv, and Y.~Bengio, ``Binarized
  neural networks,'' in \emph{Advances in neural information processing
  systems}, 2016, pp. 4107--4115.

\bibitem{DBLP:journals/corr/abs-1905-06105}
\BIBentryALTinterwordspacing
C.~Lammie, W.~Xiang, and M.~R. Azghadi, ``{Accelerating Deterministic and
  Stochastic Binarized Neural Networks on FPGAs Using OpenCL},'' \emph{CoRR},
  vol. abs/1905.06105, 2019. [Online]. Available:
  \url{http://arxiv.org/abs/1905.06105}
\BIBentrySTDinterwordspacing

\bibitem{DBLP:journals/corr/abs-1808-00737}
O.~{Krestinskaya} and A.~P. {James}, ``Binary weighted memristive analog deep
  neural network for near-sensor edge processing,'' in \emph{2018 IEEE 18th
  International Conference on Nanotechnology (IEEE-NANO)}, July 2018, pp. 1--4.

\bibitem{van2018memristor}
K.~Van~Pham, T.~Van~Nguyen, S.~B. Tran, H.~Nam, M.~J. Lee, B.~J. Choi, S.~N.
  Truong, and K.~Min, ``Memristor binarized neural networks,'' \emph{J.
  Semicond. Technol. Sci}, vol.~18, pp. 568--577, 2018.

\bibitem{8351429}
V.~{Mladenov} and S.~{Kirilov}, ``A memristor model with a modified window
  function and activation thresholds,'' in \emph{2018 IEEE International
  Symposium on Circuits and Systems (ISCAS)}, May 2018, pp. 1--5.

\end{thebibliography}

\end{document}